\documentclass[12pt]{article}
\usepackage{epsfig}
\usepackage{epstopdf}

\newcommand{\bea}{\begin{eqnarray}}
\newcommand{\eea}{\end{eqnarray}}
\newcommand{\be}{\begin{equation}}
\newcommand{\ee}{\end{equation}}

\newcommand{\as}{\alpha_s}
\newcommand{\asMZ}{\alpha_s(M^2_Z)}

\title{Improved nonsinglet QCD analysis of the fixed-target DIS data
\author{  A.V.~Kotikov, V.G.~Krivokhizhin, B.G.~Shaikhatdenov\\
Joint Institute for Nuclear Research, Russia } }

\begin{document}
\maketitle
\abstract{Deep inelastic scattering data on $F_2$ structure function obtained by BCDMS, SLAC and NMC
collaborations in fixed-target experiments are analyzed in the non-singlet
approximation with next-to-next-to-leading-order accuracy.
The strong coupling constant is found to be $\as(M_Z^2) = 0.1157 \pm 0.0022~ \mbox{\scriptsize{(total\,\,\,\,\,\,\,\,\, exp.error)}}+\biggl\{
\begin{array}{l} +0.0028 \\ -0.0016 \end{array} ~\mbox{\scriptsize{(theor)}}$, which is seen to be well compatible
with the average world value. The results, obtained in the present paper by carrying out fits similar to what were performed in [1], 
with the exception for systematic errors in BCDMS data taken into account in a different way, confirm our previous ones derived in [1].
This study is also meant to at least partially explain differences in the latest predictions for LHC observables, 
caused by usage of different sets of parton distribution functions obtained by different groups. \\

$PACS:~~12.38~Aw,\,Bx,\,Qk$\\

{\it Keywords:} Deep inelastic scattering; Nucleon structure functions;
QCD coupling constant; NNLO level; $1/Q^2$ power corrections.

\section{ Introduction }

The cross-section values obtained in LHC experiments, along with the extracted parameters,
such as, for example, the mass of $t$ quark and the strong coupling constant  $\as(M_Z^2)$,
depend crucially on the type of parton distribution funstions (PDFs) used in the analyses.
Recently, large differences are found in both the cross-section values and the extracted parameters, which 
were obtained by using Alekhin--Blumlein--Moch (ABM)~\cite{ABM1} and Jimenez-Delgado--Reya (JR) PDF sets~\cite{JR}.
The latter were in turn derived mostly by fitting deep inelastic scattering (DIS) data, which
is one of the most important processes in studying PDFs in nucleons. Other
groups doing such an analysis, namely, CTEQ~\cite{CT10}, NN21 Collaborations~\cite{NN21} and  MSTW  
group~\cite{MSTW}, included in their fits additional experimental data (see a recent review~\cite{Watt:2011kp} 
and references therein). 

The differences are sometimes seen to be much larger than the individual PDF uncertainties~\cite{Watt:2011kp,Forte:2013wc}
and give rise to mostly different shapes of gluon densities and strong coupling constant  $\as(M_Z^2)$, which are known to be strongly correlated.
The values of $\as(M_Z^2)$ obtained while using the ABM sets~\cite{ABM1,ABM} are considerably lower than those derived in other cases
and can partially be explained~\cite{Thorne:2014toa} by the usage of the fixed flavor number scheme in the ABM sets.

In the present paper we will focus on the strong coupling constant value. There is another way to decrease the value 
of $\as(M_Z^2)$  observed in~\cite{ABM1,ABM}, which is associated with a so-called {\it BCDMS effect}. 
The effect comes about upon analyzing stiffly accurate BCDMS data~\cite{BCDMS1},
which are very important in fitting the value of $\as(M_Z^2)$, especially in the analyses based on mostly DIS data,
which is the case for ABM sets. However, as it was shown in~\cite{Kri2} those precise data were collected with large systematic 
errors within certain ranges, which can presumably be responsible for an effective decrease in the value of
$\as(M_Z^2)$ (see~\cite{KKPS,Kri2,KK2001}).

One of the most accurate processes suitable for extracting $\as(M_Z^2)$ values is the valence part
of DIS structure function (SF) $F_2$, which is free of any correlations with gluon density. 
Here we will only consider the valence part\footnote{In the present paper 
we restrict analysis to the large $x$ region. Consequently, the
analysis is dubbed a ``valence quark'' or ``nonsinglet'' one  (i.e. no gluons take part in the analyses) 
but actually the data on the total structure function $F_2(x,Q^2)$ will be considered.}.
The study closely follows similar analyses~\cite{KKPS,KK2001} performed at the next- (NLO) and next-to-next-to-leading-order (NNLO) levels, respectively.
The difference between analyses done in~\cite{KKPS} and~\cite{KK2001} is that only a nonsinglet case is dealt with in the former.
In the present paper we consider systematic errors in BCDMS data in a different manner than it was done in~\cite{KKPS} in order to study how they influence
our results obtained in~\cite{KKPS,KK2001}, where $\asMZ$ value was shown to increase when we cut out BCDMS data with the largest systematic errors.
Those results have recently been questioned in~\cite{ABM}, where it was found that this effect is negligible. The authors of~\cite{ABM} conjectured that
the $\asMZ$ value increased due to neglected systematic errors in BCDMS data in the analyses done in~\cite{KKPS,KK2001}.

First, we'd like to stress that systematic errors in BCDMS data were not neglected in~\cite{KKPS,KK2001}; moreover, we are going to show that including 
those BCDMS systematic errors in a different way does not significantly modify our results derived in~\cite{KKPS,KK2001}.
Upon omitting the BCDMS data with largest systematic errors we obtain higher values of the coupling constant normalization $\asMZ$
fitted to the experimental data. Moreover, the effect does not strongly depend on specific cut values, as it was observed earlier~\cite{KKPS,KK2001}.

BCDMS, SLAC and NMC experimental data on DIS structure function (SF) $F_2(x,Q^2)$~\cite{BCDMS1,SLAC1,NMC} are analyzed at NNLO of massless perturbative QCD.
More or less reliable calculations at this level became possible due to $\alpha_s^3(Q^2)$ corrections to the splitting functions (the anomalous dimensions of Wilson operators)~\cite{MVV2004}.

\section{ Theoretical basis and fitting procedure}

As in our previous papers the function $F_2(x,Q^2)$ is represented as a sum of the leading twist $F_2^{pQCD}(x,Q^2)$ and twist four terms
\be
F_2(x,Q^2)=F_2^{pQCD}(x,Q^2)\left(1+\frac{\tilde h_4(x)}{Q^2}\right)\,.
\label{1.1}
\ee
In the analyses performed over experimental data various effects and corrections must be taken into account.
Here the nuclear effects, target mass and heavy quark threshold corrections and higher twist terms are considered. For details we refer to~\cite{KK2001,KK2009}.

There in general are two methods of carrying out a QCD analysis over DIS data: the first one (see e.g.~\cite{ABM1}-\cite{MSTW}) deals 
with Dokshitzer-Gribov-Lipatov-Altarelli-Parisi (DGLAP) integro-differential equations~\cite{DGLAP} and lets the data be examined directly,
whereas the second one involves considering SF moments and therefore allows performing an analysis in the analytic form as opposed to the former.
In the present paper we will use an approach that can be thought of as a cross between these two latter, i.e. analysis is carried out over SF moments $F_2^{k}(x,Q^2)$ defined as follows
\be
M_n^{pQCD/twist2/\ldots}(Q^2)=\int_0^1 x^{n-2}\,F_2^{pQCD/twist2/\ldots}(x,Q^2)\,dx
\label{1.a}
\ee
followed by the reconstruction of SF for each $Q^2$ by using a Jacobi polynomial expansion method~\cite{Barker,Kri} (for further details see~\cite{KK2001,KK2009}).

Aspects of the analyses related with a $Q^2$-evolution of PDF moments (with the analytic continuation~\cite{KaKo} in their coefficient functions
 and anomalous dimensions), PDF normalization, target mass (TMC) and higher twist corrections (HTCs),
as well as nuclear effects remain virtually the same as in our previous work~\cite{KK2001} so we refer to it for more details.

As usual, the moments ${\bf f}_i(n,Q^2)\,, (i = \mbox{ns, q, g})$ at some $Q^2_0$ is a theoretical input to the analysis.
In the fits of data with the cut $x\geq 0.25$ imposed, only the nonsinglet parton
density is worked with and the following patametrization at the normalization point is used (see, for example,~\cite{KKPS,PKK}):
\be
{\bf f}(n,Q^2) = \int_0^1 dx x^{n-2} \tilde{\bf f}(x,Q^2),~~~\tilde{\bf f}(x,Q^2) = 
A(Q^2)x^{\lambda(Q^2)} [1-x]^{b(Q^2)}[1+d(Q^2)x]\,,\label{4}
\ee
where $A(Q^2)$, $\lambda(Q^2)$, $b(Q^2)$ and $d(Q^2)$ are some coefficient functions.

Recall also some salient points of the so-called polynomial expansion method.
The latter was first proposed in~\cite{Ynd} and further developed in~\cite{gon}.
The Jacobi polynomials were first proposed and subsequently
developed in~\cite{Barker,Kri} with the purpose of analyzing the experimental data on SFs.
They were then extensively used in \cite{PKK}-\cite{Vovk}.

With the QCD expressions for the Mellin moments $M_n^{k}(Q^2)$ (see, for example, \cite{KK2009})
the SF $F_2^k(x,Q^2)$ is reconstructed by using the Jacobi polynomial expansion method:
$$
F_{2}^k(x,Q^2)=x^a(1-x)^b\sum_{n=0}^{N_{max}}\Theta_n ^{a,b}(x)\sum_{j=0}^{n}c_j^{(n)}(\alpha ,\beta )
M_{j+2}^k (Q^2)\,,
\label{2.1}
$$
where $\Theta_n^{a,b}$ are the Jacobi polynomials and $a,b$ are the parameters to be fitted. A condition
put on the latter is the requirement of the error minimization while reconstructing the structure functions.
MINUIT package~\cite{MINUIT} is used to minimize two variables; namely, the function $F_2$ itself and its logarithmic ``slope''
$d\ln F_2(x,Q^2)/d\ln\ln(Q^2/\Lambda^2)$.
The twist expansion is thought to be applicable above approximately $Q^2 \sim 1$ GeV$^2$ hence the global cut $Q^2 \geq 1$ GeV$^2$ imposed upon the data throughout.

The systematic errors for BCDMS data are given~\cite{BCDMS1}
as multiplicative factors to the experimental $F_2(x,Q^2)$ value. Denoted as $f_r, f_b, f_s, f_d$
and $f_h$ they are in fact uncertainties caused by the spectrometer resolution, beam momentum,
calibration, spectrometer magnetic field calibration, detector inefficiencies
and the energy normalization, respectively. In the previous paper~\cite{KKPS} the experimental value in each point of the original dataset was multiplied
by the factor characterizing the type of uncertainties, one at a time, and thus modified dataset for the analyzed quantity was then fitted by using the procedure
sketched above. Afterwards, all the systematic errors obtained this way were taken together in quadrature to yield the total one. In the present paper,
instead of using that multiplicative approach, the total systematic error for $\as$ is obtained by taking into account all those five uncertainties in quadrature
at each experimental point and then used to calculate the total experimental error.

We use free normalizations of the data for different experiments. For a reference set, the most stable deuterium BCDMS data at the value
of the beam initial energy $E_0=200$ GeV is used. When other datasets are taken as a reference one, variation in the results is found to be negligible.
In the case of the fixed normalization for each and all datasets the fits tend to yield a little bit worse $\chi^2$, just as was observed earlier.

\section{Results}

As is known a nonsinglet analysis features no gluons taking part in the analysis; therefore, the cut imposed over the Bjorken variable ($x\geq 0.25$) effectively
excludes the region where gluon density is believed to be non-negligible.

A starting point of the evolution is $Q^2_0$ = 90 GeV$^2$ for BCDMS and all datasets, and $Q^2_0$ = 20 GeV$^2$ --- for combined SLAC and NMC datasets.
These $Q^2_0$ values are close to the average values of $Q^2$ spanning the corresponding data.
The heavy quark thresholds are taken at $Q^2_f=m^2_f$.
Following our previous papers the maximum number of moments to be accounted for in the analyses is taken to be $N_{max} =8$~\cite{Kri}
and the additional cut $x \leq 0.8$ is imposed everywhere.

\subsection { $\as$ from individual BCDMS and SLAC+NMC analyses }
Analysis starts with the most precise experimental data~\cite{BCDMS1} obtained by the BCDMS muon scattering experiment for large $Q^2$ values.
A complete set of data includes 607 points when the cut $x \geq 0.25$ is imposed. 
An original analysis carried out by the BCDMS collaboration (see also~\cite{ViMi}) gave (back then) comparatively small values for the strong coupling constant;
for example, $\alpha_s(M^2_Z)=0.113$ at NLO was quoted in the latter reference.

Much like in~\cite{KKPS,KK2001} the data with largest systematic errors are cut out by imposing certain limits on the kinematic variable $Y=(E_0-E)/E_0$
(where $E_0$ and $E$ are lepton's initial and final energies, respectively~\cite{Kri2}).
The following $y$ cuts depending on the limits put on $x$ are imposed:
\bea
& &y \geq 0.14 \,~~~\mbox{ for  }~~~ 0.3 < x \leq 0.4 \nonumber \\
& &y \geq 0.16 \,~~~\mbox{ for  }~~~ 0.4 < x \leq 0.5 \nonumber \\
& &y \geq Y_{cut3} ~~~\mbox{ for  }~~~ 0.5 < x \leq 0.6 \nonumber \\
& &y \geq Y_{cut4} ~~~\mbox{ for  }~~~ 0.6 < x \leq 0.7 \nonumber \\
& &y \geq Y_{cut5} ~~~\mbox{ for  }~~~ 0.7 < x \leq 0.8 \nonumber
\eea
An impact of experimental systematic errors on the results of QCD analysis is studied for a few sets of $Y_{cut3}$, $Y_{cut4}$ and $Y_{cut5}$ cuts given in Table~1.

\newpage
{\bf Table 1.} A set of $Y_{cut3}$, $Y_{cut4}$ and $Y_{cut5}$ values used in the analysis
\footnote{This set slightly differs from that presented in~\cite{KKPS,KK2001}.}
\begin{center}
\begin{tabular}{|c|c|c|c|c|c|c|}
\hline
& & & &  & \\
$N_{Y_{cut}}$ & 1 & 2 & 3 & 4 & 5  \\
& & & &  & \\
\hline \hline
$Y_{cut3}$ &  0.16 & 0.16 & 0.18 & 0.22 & 0.23 \\
$Y_{cut4}$ &  0.18 & 0.20 & 0.20 & 0.23 & 0.24 \\
$Y_{cut5}$ & 0.20 & 0.22 & 0.22 & 0.24 & 0.25 \\
\hline
\end{tabular}
\end{center}
\vspace{0.5cm}

Following the analyses performed in~\cite{KKPS,KK2001}, we arrive at similar results: $\alpha_s$ values for both original and modified (by cuts) datasets are shown in Tables~2 and~3,
where a total systematic error is estimated in quadrature by using the method somewhat different from that utilized in our earlier analyses. ($N_{Y_{cut}}=0$ corresponds to the case without $Y$ cuts).
Namely, instead of accounting for those errors by the multiplication procedure (an old approach outlined in~\cite{KKPS}), here they are taken altogether in quadrature from the very beginning.

\vspace{0.5cm}
{\bf Table 2.} NLO $\asMZ$ values for various sets of $Y$ cuts
\begin{center}
\begin{tabular}{|l|c|c|c|c|c|}
\hline
              &           & $\chi^2$           & $\as(M_Z^2)$      & $\as(M_Z^2)$       & \\
$N_{Y_{cut}}$ & number    & quad. syst. err.   & $\pm$ stat. error & $\pm$ stat. error  & total  \\
              & of points & (mult. syst. err.) & quad. syst. err.  & mult. syst. err.   &  syst. error \\
\hline \hline
0 & 607 & 444 (609) & 0.1078 $\pm$ 0.0012 & 0.1072 $\pm$ 0.0012 & 0.0054\\
1 & 502 & 358 (477) & 0.1149 $\pm$ 0.0015 & 0.1146 $\pm$ 0.0015 & 0.0039\\
2 & 495 & 355 (469) & 0.1151 $\pm$ 0.0015 & 0.1148 $\pm$ 0.0015 & 0.0038\\
3 & 489 & 350 (459) & 0.157 $\pm$ 0.0015 & 0.1155 $\pm$ 0.0015 & 0.0036\\
4 & 458 & 327 (423) & 0.1166 $\pm$ 0.0016 & 0.1163 $\pm$ 0.0016 & 0.0031\\
5 & 452 & 322 (417) & 0.1172 $\pm$ 0.0016 & 0.1168 $\pm$ 0.0016 & 0.0030\\
\hline
\end{tabular}
\end{center}

\vspace{0.5cm}
{\bf Table 3.} NNLO $\asMZ$ values for various sets of $Y$ cuts
\begin{center}
\begin{tabular}{|l|c|c|c|c|c|}
\hline
              &           & $\chi^2$           & $\as(M_Z^2)$      & $\as(M_Z^2)$      & \\
$N_{Y_{cut}}$ & number    & quad. syst. err.   & $\pm$ stat. error & $\pm$ stat. error & total  \\
              & of points & (mult. syst. err.) & quad. syst. err.  & mult. syst. err.  &  syst. error \\
\hline \hline
0 & 607 & 446 (642) & 0.1064 $\pm$ 0.0012 & 0.1056 $\pm$ 0.0012 & 0.0054\\
1 & 502 & 361 (481) & 0.1132 $\pm$ 0.0015 & 0.1127 $\pm$ 0.0015 & 0.0039\\
2 & 495 & 357 (477) & 0.1135 $\pm$ 0.0015 & 0.1130 $\pm$ 0.0015 & 0.0038\\
3 & 489 & 352 (463) & 0.1140 $\pm$ 0.0015 & 0.1136 $\pm$ 0.0015 & 0.0036\\
4 & 458 & 350 (427) & 0.1150 $\pm$ 0.0016 & 0.1144 $\pm$ 0.0016 & 0.0031\\
5 & 452 & 325 (421) & 0.1155 $\pm$ 0.0016 & 0.1149 $\pm$ 0.0016 & 0.0030\\
\hline
\end{tabular}
\end{center}
\vspace{0.5cm}

Once the cuts are imposed (in what follows we work with a set $N_{Y_{cut}}=5$), only 452 points left available for analysis.
Fitting them according to the procedure outlined above the following results are obtained:
\bea
\as(M_Z^2) = 0.1155 +
\biggl\{ \begin{array}{l}
\pm 0.0016~\mbox{(stat)}
\pm 0.0030~\mbox{(syst)} \pm 0.0007~\mbox{(norm)}\\
\pm 0.0035~\mbox{(total exp. error)}
\end{array} \label{bd1a}\,,
\eea
where an abbreviation ``norm'' denotes the experimental data normalization error stemming from the difference of the fits
with free and fixed normalizations of BCDMS data subsets~\cite{BCDMS1} featuring different beam energy values.

In view of a few NMC data points in the region $x \geq 0.25$, we have performed similar combined fits of SLAC and NMC datasets
and found the following NNLO result for the strong coupling normalization (which is obtained under the same conditions 
as the one presented in the last row of Table~IV in~\cite{KKPS}):
\bea
\as(M_Z^2) = 0.1180 +
\biggl\{
\begin{array}{l}
\pm 0.0014~\mbox{(stat)}
\pm 0.0036~\mbox{(syst)} \pm 0.0008~\mbox{(norm)}\\
\pm 0.0039~\mbox{(total exp. error)}\,.
\end{array} 
\eea

\subsection { $\as$ and HTC parameters in combined SLAC, BCDMS and NMC analysis }
As in the case of BCDMS data analysis the cuts imposed are $x\geq 0.25$ and $N_{Y_{cut}}=5$ (see Table~1). Then, an overall set of data consists of 756 points.

In order to determine the region where perturbative QCD is applicable we start by analyzing the data without a contribution of twist-four terms
(that is $F_2 = F_2^{pQCD}$) and perform several fits with the cut $Q^2 \geq Q^2_{min}$ gradually increased.
Table~4 demonstrates that the quality of fits appears to be already acceptable beginning from $Q^2=2$ GeV$^2$.

Now, the twist-four corrections are added and the data with a global cut $Q^2 \geq 1$ GeV$^2$ is fitted. As in the previous studies~\cite{KKPS,KK2001} it is clearly seen that
higher twists improve the fit quality, with an insignificant discrepancy in the values of the coupling constant to be quoted below.

\vspace{0.5cm}
{\bf Table 4.} $\asMZ$ {\sl and} $\chi^2$ {\sl in the combined SLAC, BCDMS, NMC analysis}
\begin{center}
\begin{tabular}{|l|c|c|c|c|c|c|}
\hline
& &  & &  &\\
$Q^2_{min}$ & $N$ of & HTC &$\chi^2(F_2)$/DOF &
$\as(90~\mbox{GeV}^2)$ $\pm$ stat & $\asMZ$ \\
& points &  &  & &  \\
\hline \hline
1.0 & 756 &  No & 1.41 & 0.1757 $\pm$ 0.0007 & 0.1160 \\
2.0 & 731 &  No & 1.03 & 0.1758 $\pm$ 0.0007 & 0.1161 \\
3.0 & 704 &  No & 0.84 & 0.1787 $\pm$ 0.0009 & 0.1173 \\
4.0 & 682 &  No & 0.79 & 0.1790 $\pm$ 0.0009 & 0.1174 \\
5.0 & 662 &  No & 0.79 & 0.1795 $\pm$ 0.0011 & 0.1177 \\
6.0 & 637 &  No & 0.79 & 0.1798 $\pm$ 0.0013 & 0.1178 \\
7.0 & 610 &  No & 0.78 & 0.1792 $\pm$ 0.0016 & 0.1175 \\
8.0 & 594 &  No & 0.79 & 0.1787 $\pm$ 0.0019 & 0.1173 \\
9.0 & 575 &  No & 0.78 & 0.1785 $\pm$ 0.0023 & 0.1172 \\
10.0 & 564 & No & 0.77 & 0.1765 $\pm$ 0.0026 & 0.1164 \\
\hline \hline
1.0 & 756 & Yes & 0.88 & 0.1750 $\pm$ 0.0019 & 0.1157 \\
\hline
\end{tabular}
\end{center}
\vspace{0.5cm}

The following values for PDF parametrization parameters are obtained in the fit corresponding to the last row of the above table
 (errors shown are actually the symmetric ones):
\bea
A(H_2) &=& 3.443 \pm 0.046,~~~\, A(D_2) ~=~ 9.201\pm 0.111,~~~\, A(C) ~=~ 7.580 \pm 0.173,\nonumber \\
\lambda(H_2)  &=& 0.143 \pm 0.007,~~~\, \lambda(D_2) ~=~ 0.494 \pm 0.007,~~~\,  \lambda(C) ~=~ 0.340 \pm 0.015,\nonumber \\
b(H_2) &=& 4.144 \pm 0.013,~~~\, b(D_2) ~=~ 3.795 \pm 0.019,~~~\, b(C) ~=~ 3.615  \pm 0.036,\nonumber \\
d(H_2) &=& 3.223\pm 0.072,~~~\,  d(D_2) ~=~ -0.404\pm 0.024,~~~\,  d(C) ~=~-0.629 \pm 0.034.\nonumber
\label{Norma}
\eea
They are seen to be very similar to those presented in~\cite{KKPS}.

Twist-four parameter values are presented in Table~5.
Note that these for $H_2$ and $D_2$ targets are obtained in separate fits by analyzing SLAC, NMC and BCDMS datasets altogether.

\vskip 0.5cm
{\bf Table 5.} {\sl Parameter values of the twist-four term in different orders}
\begin{center}
\scriptsize
\begin{tabular}{|c||c|c||c|c||c|c|}
\hline
 & \multicolumn{2}{c||}{LO $\pm$ stat} & \multicolumn{2}{c||}{NLO $\pm$ stat} & \multicolumn{2}{c|}{NNLO $\pm$ stat}  \\
\cline{2-7}
$x$ & $\tilde h_4(x)$ for $H_2$ & $\tilde h_4(x)$ for $D_2$ & $\tilde h_4(x)$ for $H_2$ & $\tilde h_4(x)$ for $D_2$ & $\tilde h_4(x)$ for $H_2$ & $\tilde h_4(x)$ for $D_2$ \\
\hline \hline
0.275&-0.275$\pm$0.012&-0.274$\pm$0.021&-0.251$\pm$0.026&-0.221$\pm$0.007& -0.176 $\pm$ 0.023  & -0.136 $\pm$ 0.026\\
0.35 &-0.269$\pm$0.015&-0.246$\pm$0.030&-0.190$\pm$0.025&-0.178$\pm$0.006& -0.143 $\pm$ 0.032  & -0.021 $\pm$ 0.015 \\
0.45 &-0.181$\pm$0.016&-0.125$\pm$0.049&-0.290$\pm$0.017&-0.002$\pm$0.010& -0.181 $\pm$ 0.031  & -0.068 $\pm$ 0.015 \\
0.55 &-0.049$\pm$0.033&0.080 $\pm$0.082&-0.316$\pm$0.031&0.247 $\pm$0.014& -0.233 $\pm$ 0.059  &  0.043 $\pm$ 0.022 \\
0.65 &0.326 $\pm$0.063&0.357 $\pm$0.144&-0.064$\pm$0.076&0.563 $\pm$0.035& -0.167 $\pm$ 0.156  &  0.353 $\pm$ 0.059 \\
0.75 &0.805 $\pm$0.124&0.513 $\pm$0.213&0.009 $\pm$0.122&0.770 $\pm$0.070& -0.156 $\pm$ 0.215  &  0.323 $\pm$ 0.104 \\
\hline
\end{tabular}
\end{center}
\vspace{0.5cm}

\begin{figure}[!htb]
\unitlength=1mm
\vskip -1.5cm
\begin{picture}(0,100)
  \put(0,-5){%
   \psfig{file=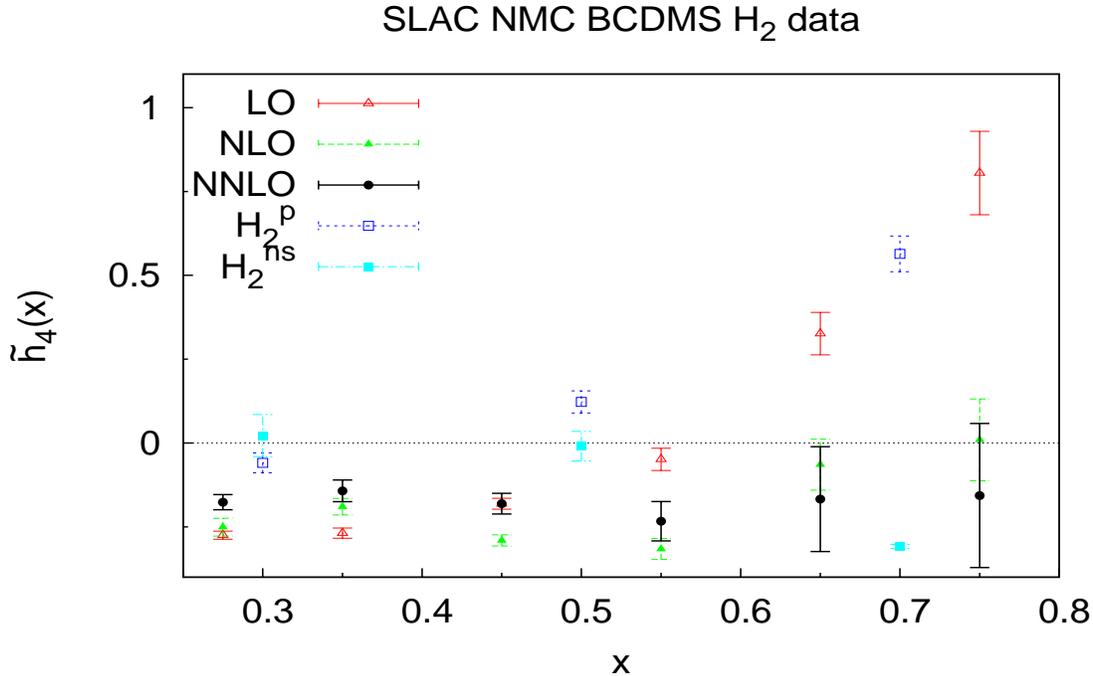,width=150mm,height=95mm}
}
\end{picture}
\vskip 0.2cm
\caption{ Twist-four $\tilde h_4(x)$ parameter values obtained at LO, NLO and NNLO for hydrogen data (bars show statistical errors).
For comparison, H$_2^p$ and H$_2^{ns}$ points borrowed from~\cite{ABM1} are also shown.}
\end{figure}

\begin{figure}[htb]
\unitlength=1mm
\vskip -1.5cm
\begin{picture}(0,100)
\put(0,-5){\psfig{file=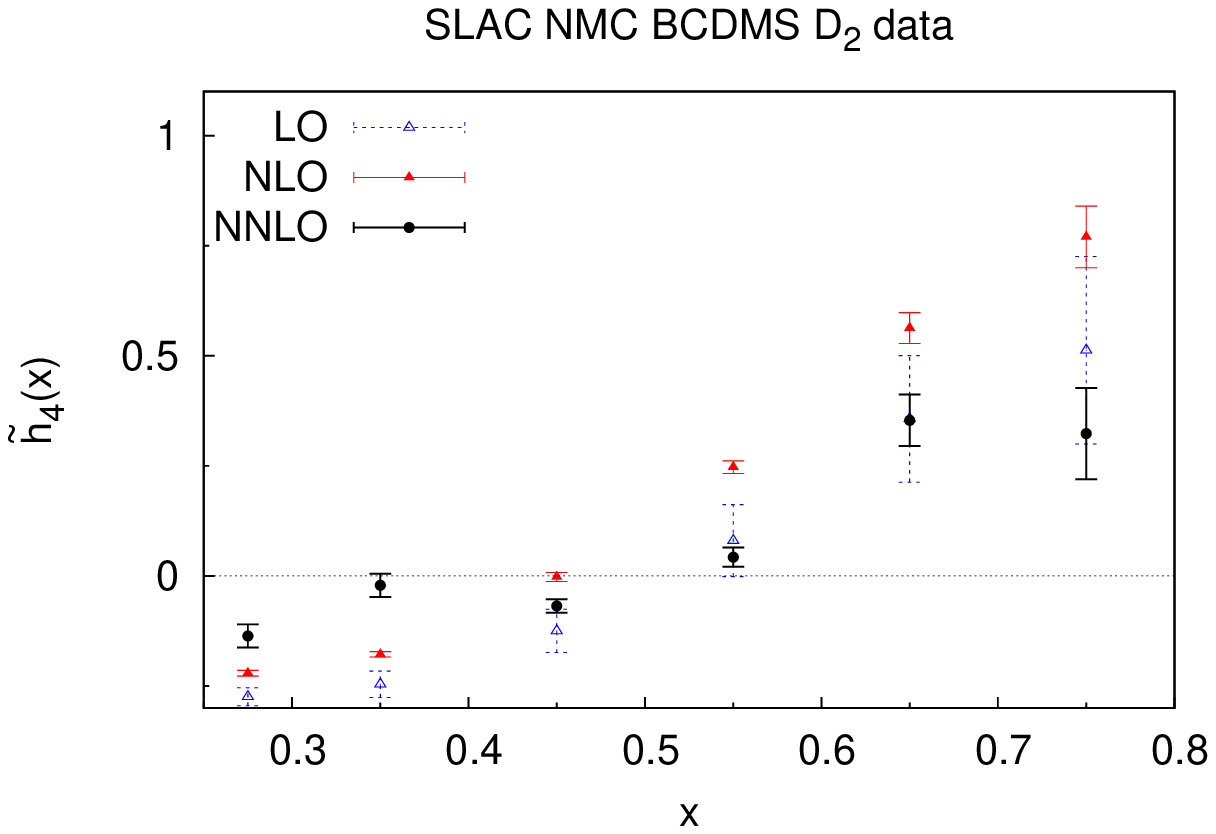,width=150mm,height=95mm}}
\end{picture}
\vskip 0.2cm
\caption{ The same as in Fig.~1 for deuterium data.}
\end{figure}
\vspace{0.5cm}

HTCs are shown in Figs.~1 and~2, where twist-four corrections obtained at NLO and NNLO are observed to be compatible with each other. 
Their values in Table~5 are very similar to those presented in~\cite{KKPS}. Therefore, all the comments in~\cite{KKPS}
can be applied also to the present case and we wont be repeating discussions given there. We would only like to stress
that the cut imposed on BCDMS data, which effectively increased $\as$ values (see Tables~2 and~3), essentially 
improves also an agreement between theory and experiment.
HTCs as being the difference between the twist-two approximation (i.e. pure perturbative QCD contribution) and the experimental data becomes considerably smaller at NLO and NNLO levels
compared with NLO HT terms obtained in~\cite{ViMi} and also with the results of analysis with no $Y$-cuts imposed over BCDMS data (see Figs.~5,~6 in~\cite{KKPS}).

Some additional support of these observations can be gained by considering NNLO HTCs presented in~\cite{ABM}, which are shown 
\footnote{The authors of~\cite{ABM} used another HTC definition. Upon translating the latter to our definition it looks like
$F_2(x,Q^2)=F_2^{QCD}(x,Q^2) +\tilde H_4(x)/Q^2$. So, the functions $\tilde h_4(x)$ can be estimated in the following way:
$\tilde h_4(x)=\tilde H_4(x)/F_2^{QCD}(x,Q^2)$, where we can use $Q^2=2.5$ GeV$^2$, which is a lower $Q^2$ 
boundary in the analysis~\cite{ABM}.}
in Fig. 1. HTCs obtained in~\cite{ABM} are seen to be virtually higher than our ones. However, they are well compatible with the corresponding HTCs,
obtained in the analysis with no $Y$-cuts imposed over BCDMS data (see Fig.~5 in~\cite{KKPS}).

Figure 1 also displays HTC results for the nonsinglet part of $F_2$ borrowed from~\cite{ABM}. They demonstrate the difference 
in the HTC shapes for $H_2$ and $D_2$ cases, which agrees with our results presented in Table 5 and Figs. 1 and 2.

In order to assess the quality of fits we present the correlation matrix of the fit parameters (Appendix~A) and pulls for individual datasets (Appendix~B).

Finally, using the nonsinglet evolution analyses of SLAC, NMC and BCDMS experimental data for SF $F_2$ with no account for twist-four corrections and the cut $Q^2 \geq 2$ GeV$^2$,
we obtain (with $\chi^2/DOF=1.03$)
\bea
\as(M_Z^2) ~=~ 0.1161 +
\biggl\{ \begin{array}{l} \pm 0.0003 ~\mbox{(stat)}
\pm 0.0018~\mbox{(syst)}  \pm 0.0007 ~\mbox{(norm)}\\
\pm 0.0020~\mbox{(total exp.error)}
 \end{array} \,.
\eea

Upon including the twist-four corrections, and imposing the cut $Q^2 \geq 1$ GeV$^2$, the following result is found (with $\chi^2/DOF=0.88$):
\bea
\as(M_Z^2) ~=~ 0.1157 +
\biggl\{ \begin{array}{l}
\pm 0.0008 ~\mbox{(stat)}
\pm 0.0020 ~\mbox{(syst)}  \pm 0.0005 ~\mbox{(norm)} \\
\pm 0.0022 ~\mbox{(total exp.error)}
\end{array} \,.
\eea

Results for $n=2$ moment of the difference of valence parts of the $u$ and $d$ quarks are also investigated
in the lattice models. Following~\cite{NNLOBlumlein,ABM}, we will try to estimate this second moment.

They can be extracted at large $x$ values directly from the difference of the nonsiglet parton densities in proton and deutron:
\be
\tilde{\bf f}_u^v(x,Q^2) - \tilde{\bf f}_d^v(x,Q^2) \approx  \tilde{\bf f}_p(x,Q^2)-\tilde{\bf f}_d(x,Q^2),
\ee
because in this case the contribution of sea quarks and antiquarks is negligible (see, for example, recent papers~\cite{Alekhin:2014sya}).

This way, by using the results given in Eq.~(\ref{Norma}) we obtain the difference of second moments ${\bf f}_u^v(2,Q^2)-{\bf f}_d^v(2,Q^2)$,
shown in Table 6.
\begin{figure}[!htb]
\unitlength=1mm
\vskip -1.5cm
\begin{picture}(0,100)
  \put(0,-5){%
   \psfig{file=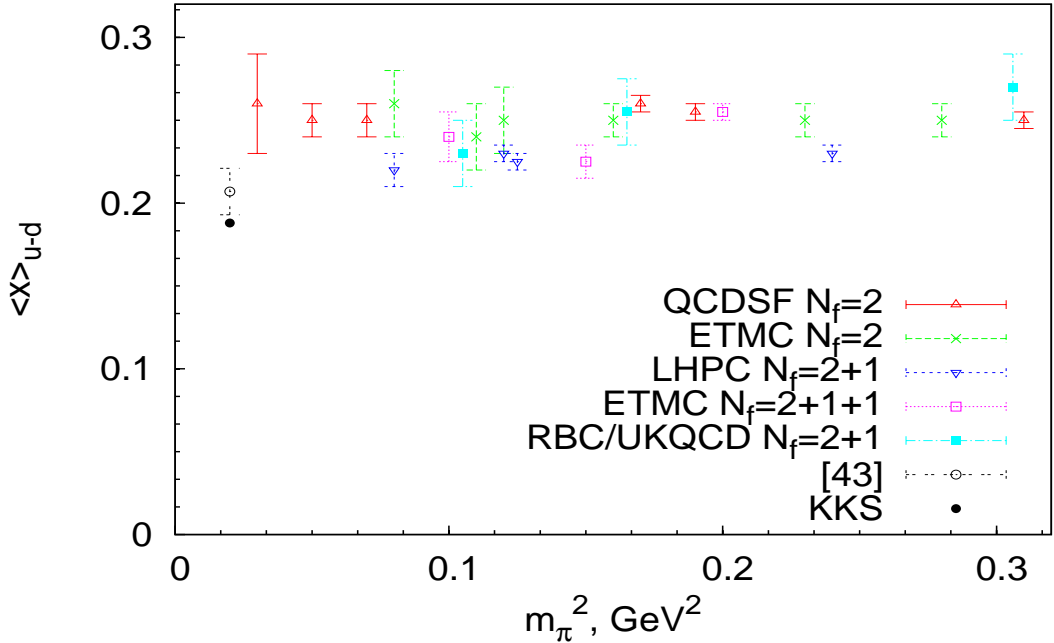,width=150mm,height=95mm}
}
\end{picture}
\vskip 0.2cm
\caption{ Results of lattice computations for the second moment of non-singlet density
as a function of the pion mass $m_{\pi}$. Our result (KKS) for 
$<x>_{u-d} \equiv {\bf f}_u^v(2,Q^2) - {\bf f}_d^v(2,Q^2)$ is given for $Q^2$ = 1 GeV$^2$.}
\end{figure}

{\bf Table 6.} Results for the difference ${\bf f}_u^v(2,Q^2)-{\bf f}_d^v(2,Q^2)$.
\vspace{0.2cm}
\begin{center}
\begin{tabular}{|c|c|c|c|c|}
\hline
$Q^2$, GeV$^2$ & 90     & 4       & 2                             & 1 \\ \hline \hline
diff.   & 0.115 $\pm$ 0.053 & 0.146 $\pm$ 0.067 & 0.157 $\pm$ 0.073 & 0.188 $\pm$ 0.087  \\ \hline
\end{tabular}
\end{center}

\vskip 0.5cm
As seen from Table 6, at $Q^2$= 4 GeV$^2$ the difference derived in the present paper is approximately $10\div15\%$ less than that
quoted in~\cite{ABM,NNLOBlumlein}. Note that the lattice results are strongly nonperturbative; it would therefore
be better to compare them with our result obtained at $Q^2$= 1 GeV$^2$, which corresponds in the analysis under consideration
to the boundary between perturbative and nonperturbative QCD. This comparison with the respective lattice results is presented in Fig.~3.
Being a difference of two large quantities it has a large error and therefore only a central value of $<x>_{u-d}$ is shown in the figure.

The  lattice points except for the leftmost one are taken from the lattice collaborations: QCDSF ($n_f=2$) 
\cite{Gockeler:2011se}, RBC/UKQCD ($n_f=2+1$) \cite{Aoki:2010xg}, LHPC ($n_f=2+1$) \cite{Bratt:2010jn}, ETMC ($n_f=2$)
\cite{Alexandrou:2011nr},  ETMC  ($n_f=2+1$) \cite{Dinter:2011jt,Dinter:2011sg}. The  leftmost point is borrowed from the
recent paper~\cite{Bali:2012av}. As seen in Fig.~3, it agrees well with our result for $ {\bf f}_u^v(2,Q^2) - {\bf f}_d^v(2,Q^2)$ 
obtained at $Q^2$= 1 GeV$^2$. Note that the result similar to those published in~\cite{Bali:2012av}, have recently been 
obtained in~\cite{Green:2012ud} (see review in~\cite{Alexandrou:2014yha}).

\section{Scale dependence}

In this section the dependence of the results on the different choice of the factorization $\mu_F=k_F Q^2$ and renormalization $\mu_R=k_R Q^2$ scales
is studied. The threshold crossing point is taken at $Q^2_f=m^2_f$.
Following~\cite{ViMi} we choose three values ($1/2,~1,~2$) for the coefficients $k_F$ and $k_R$.

Results are demonstrated in Table~7. Fits are performed with no account for higher twist corrections, the number of points is $731$ (SLAC, BCDMS and NMC data), 
$Q^2_{min} = 2$ GeV$^2$ and different datasets are freely normalized. The change in $\asMZ$ value for various $k_F$ and $k_R$ values is denoted by the difference:
\bea
\Delta \as(M_Z^2) ~=~ \as(M_Z^2) - \as(M_Z^2)|_{k_F=k_R=1}
\eea

\vspace{0.5cm}
{\bf Table 7.} $\asMZ$ {\sl for a set of} $k_F$ and $k_R$ {\sl coefficients}
\vspace{0.2cm}
\begin{center}
\begin{tabular}{|c|c||c|c|c|c|}
\hline
& & & & & \\
$k_R$ & $k_F$ & $\chi^2(F_2)$ & $\as(90~\mbox{GeV}^2)$  $\pm$ stat & $\asMZ$ & $\Delta \as(M_Z^2)$ \\
& & & & & \\ \hline \hline
1   &  1   & 712 & 0.1758 $\pm$ 0.0007   & 0.1161   & 0 \\
1/2 &  1   & 660 & 0.1729 $\pm$ 0.0007   & 0.1149   & -0.0012 \\
1   &  1/2 & 587 & 0.1733 $\pm$ 0.0006   & 0.1150   & -0.0011 \\
1   &  2   & 852 & 0.1808 $\pm$ 0.0008   & 0.1182   & +0.0021 \\
2   &  1   & 785 & 0.1805 $\pm$ 0.0008   & 0.1180   & +0.0019 \\
\hline
\end{tabular}
\end{center}
\vspace{0.5cm}

As seen from this table theoretical uncertainties for the maximum and minimum values of the coupling constant corresponding to $k_i=2$ and $k_i=1/2$ ($i=F,R$), respectively,
are found to be $+0.0028$ and $-0.0016$.\footnote{
Note that here we take theoretical errors for the factorization and renormalization scales in quadrature. In our previous analyses~\cite{KKPS,KK2001}
we considered the cases with $k_F=k_R=1/2$ and $k_F=k_R=2$ that corresponded to taking the scales together linearly rather than in quadrature.}
It should be noted that we take into account renormalization scale uncertainty in the expressions for the coefficient functions and respective coupling constants
analogously to what was done in~\cite{NeVo}.

Thus, NS evolution analyses of SLAC, NMC and BCDMS experimental data for SF $F_2$ give for $\asMZ$ the following numbers (with no account for HTC, $Q^2 \geq 2$ GeV$^2$ and $\chi^2=1.03$):
\bea
\as(M_Z^2) = 0.1161 +
\biggl\{\begin{array}{l}
\pm 0.0003 ~\mbox{(stat)}
\pm 0.0018 ~\mbox{(syst)}  \pm 0.0007 ~\mbox{(norm)} \\
\pm 0.0020 ~\mbox{(total exp.error)} \end{array}
+ \biggl\{
\begin{array}{l}+ 0.0028 \\ -0.0016 \end{array} ~\mbox{(theor)}\,.
\nonumber
\eea

\section{Conclusions}

In this work the Jacobi polynomial expansion method developed in~\cite{Barker, Kri} was used to analyze $Q^2$-evolution of DIS structure function $F_2$
by fitting reliable fixed-target experimental data that satisfy the cut $x \geq 0.25$.
Based on the results of fitting, the strong coupling constant value is evaluated at the normalization point.
Starting with the reanalysis of BCDMS data by cutting off the points with largest systematic errors it is shown that as earlier $\asMZ$ values rise sharply with the cuts on systematics imposed.
On the other hand, the latter do not depend on a certain cut within statistical errors.
The present results are compatible with those obatined in our earlier paper~\cite{KKPS}, where systematic errors in BCDMS data were taken into account in a different way.
To be more precise, in~\cite{KKPS}, and in even earlier studies~\cite{Kri,KK2001}, systematics was dealt with as follows:
all fits were done with experimental data multiplied by respective systematical errors for $F_2$ separately for each source of uncertainties.
Then, the differences between fits with different sources taken into account give rise to the total systematic error derived in quadrature. In the present paper, BCDMS systematics is dealt 
with in a quadrature rather than in a multiplicative manner right from the start.

Taking into account systematic errors in BCDMS data does not change results of the fits obtained in~\cite{KKPS} except for just
a single detail:
now perturbative QCD (without HTC) is well compatible with the experimental data already at $Q^2 \geq 2$ GeV$^2$
(see Table 4).

Here we would like to give some explanations of no-rise effect in $\asMZ$ value upon cutting out the regions in BCDMS data with larger
systematic errors stated in~\cite{ABM}. Note that the values of systematic errors are rather large in the cut out regions but not infinitely large.
In the latter case there of course is no any effect of absence/existence of the cut out regions.
One of possible explanations relates with the fact that Ref.~\cite{ABM} includes combined singlet and nonsinglet analyses, where there is some correlation between $\asMZ$ values
and the shape of gluon density. So, cutting out BCDMS data with the largest systematic errors could lead
in~\cite{ABM} to the shape of gluon density somewhat altered.

It turns out that for $Q^2 \geq 2$ GeV$^2$ the formul\ae\, of pure perturbative QCD (i.e. twist-two approximation along with the target mass corrections)
are enough to achieve good agreement with all the data analyzed. The reference result is then found to be
\be
\as(M_Z^2) = 0.1161 \pm 0.0020~\mbox{(total exp.error)}, \label{re1n} \\
\ee

Upon adding twist-four corrections, fairly good agreement between QCD and the data starting already at $Q^2 = 1$ GeV$^2$, where the Wilson
expansion starts to be applicable, is observed. This way we obtain for the coupling constant at $Z$ mass peak:
\be
\as(M_Z^2) = 0.1157 \pm 0.0022~\mbox{(total exp.error)} \, . \label{re2n} \\
\ee

Note that in a sense, our results are between those obtained in~\cite{ABM,ABM1} and~\cite{JR} (a dynamical approach) and, respectively, results derived by MSTW~\cite{MSTW}
and  NN21~\cite{NN21} groups. They are consistent with those obtained in~\cite{CT10},~\cite{JR} (a standard fit) and also with studies
of the recent data of CMS and ATLAS collaborations done in~\cite{CMS} and~\cite{ATLAS}, respectively
(see the recent review ~\cite{Rojo}).
A complete agreement with recent results obtained in lattice QCD~\cite{lattice} is also observed.
There also is very good agreement with lattice QCD results for the second moment $<x>_{u-d}$ (see Fig. 3).
Our result is slightly below the central world average value
\be
\as(M_Z^2)|_{\rm{world~average}} = 0.1185 \pm 0.0006 \, ,\label{world} \\
\ee
presented in~\cite{Breview}, but still compatible within errors.

As was already pointed out the values of theoretical uncertainties, given by this dependence of the results for $\as(M_Z^2)$ are equal to
\bea
\Delta\as(M_Z^2)|_{\mbox{theor}} ~=~
\biggl\{ \begin{array}{l} +0.0028 \\
-0.0016 \end{array} \,.
\eea

It is seen that the theoretical uncertainties are already comparable with the total experimental error. Nontheless, further account of even higher corrections is still desirable
and the next step is to consider further corrections (i.e. those coming from three loops) in the coefficient functions~\cite{MVV2005}, which allows performing N$^3$LO fits at large $x$ values,
where the contributions of the corresponding four-loop corrections to yet unknown anomalous dimensions should be negligible.
Note that several N$^3$LO fits had already been done in~\cite{KPS1,NNLOBlumlein}.

The results obtained in the present paper might be useful in shedding some extra light on the differences in the predictions
for observables at the LHC found recently~\cite{Watt:2011kp,Forte:2013wc}, which are resulted
from the utilization of different sets of parton distribution functions obtained by different groups. 
Indeed, excluding the ranges with largest systematic errors in BCDMS data increases the value of $\as(M_Z^2)$ in the fits 
based on mostly DIS experimental data and could therefore potentially lead to those differences somewhat reduced.

In order to check whether there is effect of the rise in $\asMZ$ value when $y$-cuts are imposed, it is needed to consider combined nonsinglet and singlet analysis
of DIS experimental data over an entire $x$ region. An application of some resummations, like a Grunberg's effective charge method~\cite{Grunberg}
(as it was done in~\cite{Vovk} at the NLO approximation) and the ``frozen'' (see~\cite{Zotov} and references therein)
and analytic~\cite{SoShi} versions of the strong coupling constant (see~\cite{Zotov,CIKK09,ShiTer} for recent studies in this direction) could also be useful 
in understanding the subject. These are left for the future investigations.

\section{Acknowledgments}
The work was supported by RFBR grant No.13-02-01005-a.

\section{Appendix A}
Here we present a correlation matrix of the fit parameters derived in the case shown in the last row of Table~4.
As it is seen, there are non-negligible correlations between carbon $A, b, d$ and hydrogen $d$ and deuterium $A, d$ parameters, while no appreciable correlations
are observed for the strong coupling and Jacobi parameters. By varying fit parameters and their limits we have verified that the final results are not significantly affected
by these correlations, though the latter were more or less persistent.

\vspace{0.5cm}
{\bf Table~8.} {\sl Correlation matrix of the parameters for the combined fit with HTC}
\vspace{0.2cm}
\begin{center}
\begin{tabular}{|c||c|c|c|c|c|c|c|c|c|c|c|c|}
\hline
& $A_{H_2}\!\!\!\!$ & $b_{H_2}\!\!\!\!$ & $d_{H_2}\!\!\!\!$ & $A_{D_2}\!\!\!\!$ & $b_{D_2}\!\!\!\!$ & $d_{D_2}\!\!\!\!$ & $\as(Q_0^2)\!\!\!\!$ & $A_C\!\!\!\!$ & $b_C\!\!\!\!$ & $d_C\!\!\!\!$ & b & a
 \\ \hline \hline
$A_{H_2}$   &  1.000   &         &         &         &         &         &            &       &       &       &       &         \\
$b_{H_2}$   &  0.271   &  1.000  &         &         &         &         &            &       &       &       &       &         \\
$d_{H_2}$   & -0.721   &  0.298  &  1.000  &         &         &         &            &       &       &       &       &         \\
$A_{D_2}$   & -0.593   &  0.303  &  0.873  &  1.000  &         &         &            &       &       &       &       &         \\
$b_{D_2}$   &  0.734   &  0.011  & -0.743  & -0.491  &  1.000  &         &            &       &       &       &       &         \\
$d_{D_2}$   &  0.719   & -0.304  & -0.963  & -0.893  &  0.734  &  1.000  &            &       &       &       &       &         \\
$\as(Q_0^2)$&  0.019   & -0.278  & -0.387  & -0.544  & -0.017  &  0.339  &   1.000    &       &       &       &       &         \\
$A_C    $   &  0.672   & -0.306  & -0.921  & -0.917  &  0.670  &  0.955  &   0.359    & 1.000 &       &       &       &         \\
$b_C    $   & -0.727   &  0.249  &  0.955  &  0.896  & -0.716  & -0.961  &  -0.407    &-0.891 & 1.000 &       &       &         \\
$d_C    $   & -0.716   &  0.290  &  0.964  &  0.936  & -0.710  & -0.987  &  -0.396    &-0.972 & 0.967 & 1.000 &       &         \\
b           &  0.164   &  0.369  & -0.027  &  0.131  &  0.209  &  0.003  &   0.144    & 0.006 & 0.045 & 0.017 & 1.000 &         \\
a           & -0.347   & -0.147  &  0.193  &  0.000  & -0.477  & -0.202  &   0.552    &-0.165 & 0.164 & 0.167 & 0.286 &  1.000  \\
\hline
\end{tabular}
\end{center}

\newpage
\section{Appendix B. Pulls for the individual datasets}

\begin{figure}[!htb]
\unitlength=1mm
\vskip -1.5cm
\begin{picture}(0,100)
  \put(0,-5){%
   \psfig{file=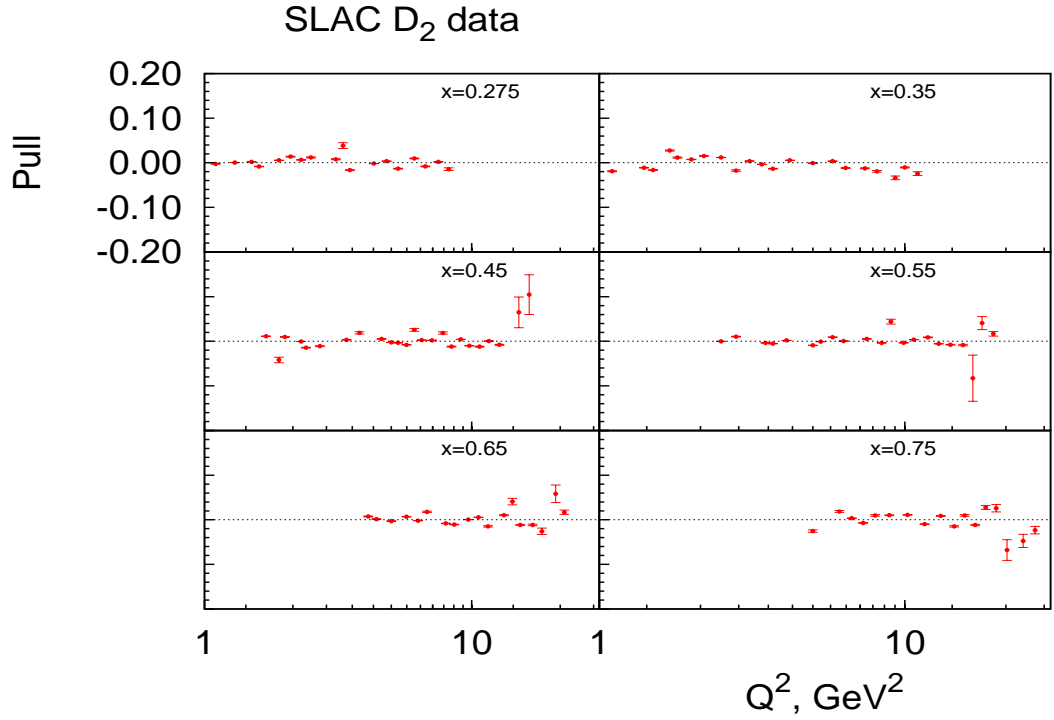,width=150mm,height=95mm}
}
\end{picture}
\vskip 0.2cm
\caption{ Pulls obtained in NNLO QCD analysis of SLAC deuterium data (bars include statistical and systematic uncertainties).}
\end{figure}

\begin{figure}[!htb]
\unitlength=1mm
\vskip -1.5cm
\begin{picture}(0,100)
  \put(0,-5){%
   \psfig{file=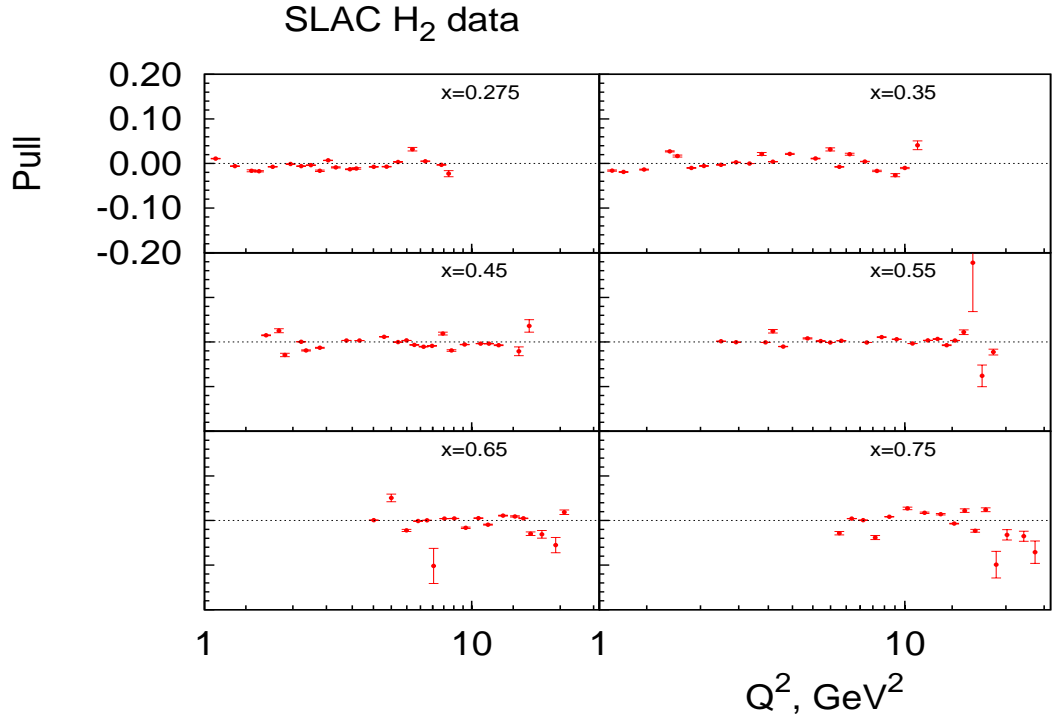,width=150mm,height=95mm}
}
\end{picture}
\vskip 0.2cm
\caption{ The same as in Fig.~1 for SLAC hydrogen data.}
\end{figure}

\vspace{1cm}
\begin{figure}[!htb]
\unitlength=1mm
\begin{picture}(0,100)
  \put(0,-5){%
   \psfig{file=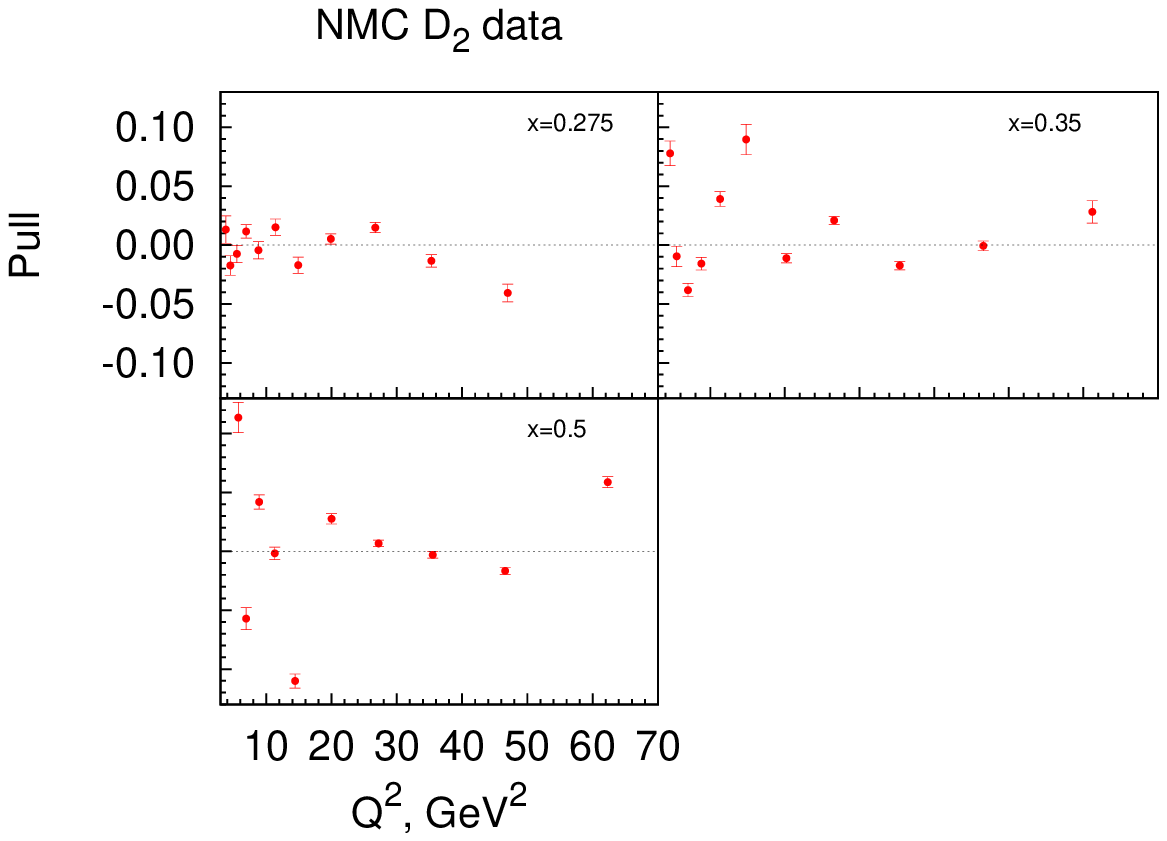,width=150mm,height=95mm}
}
\end{picture}
\vskip 0.2cm
\caption{ The same as in Fig.~1 for NMC deuterium data.}
\end{figure}

\vspace{1cm}
\begin{figure}[!htb]
\unitlength=1mm
\begin{picture}(0,100)
  \put(0,-5){%
   \psfig{file=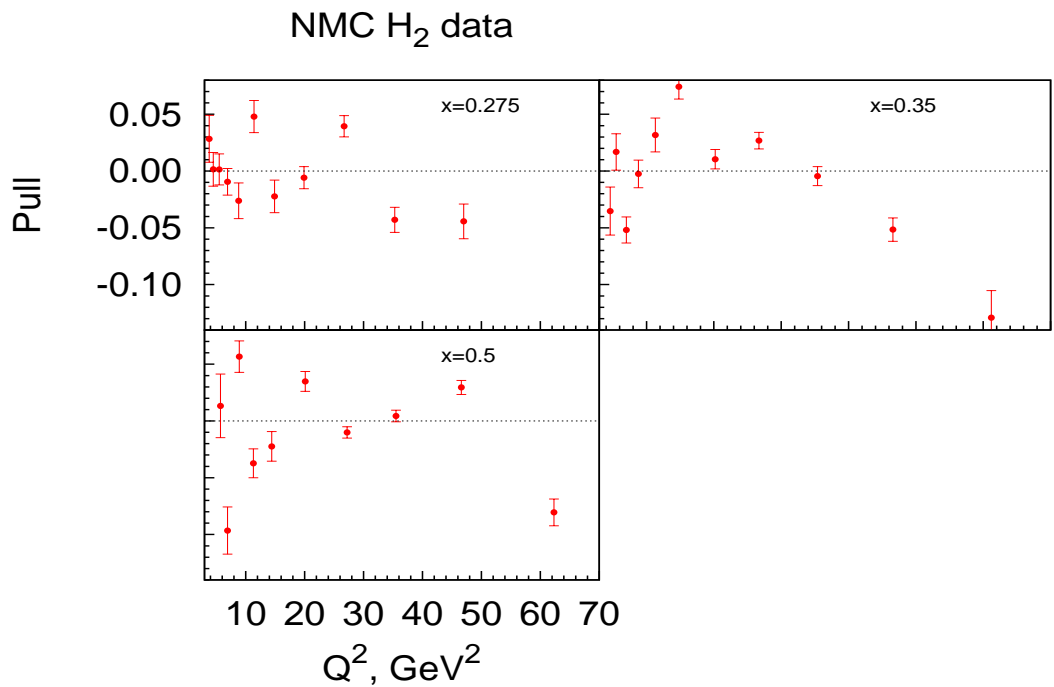,width=150mm,height=95mm}
}
\end{picture}
\vskip 0.2cm
\caption{ The same as in Fig.~1 for NMC hydrogen data.}
\end{figure}

\begin{figure}[!htb]
\unitlength=1mm
\begin{picture}(0,100)
  \put(0,-5){%
   \psfig{file=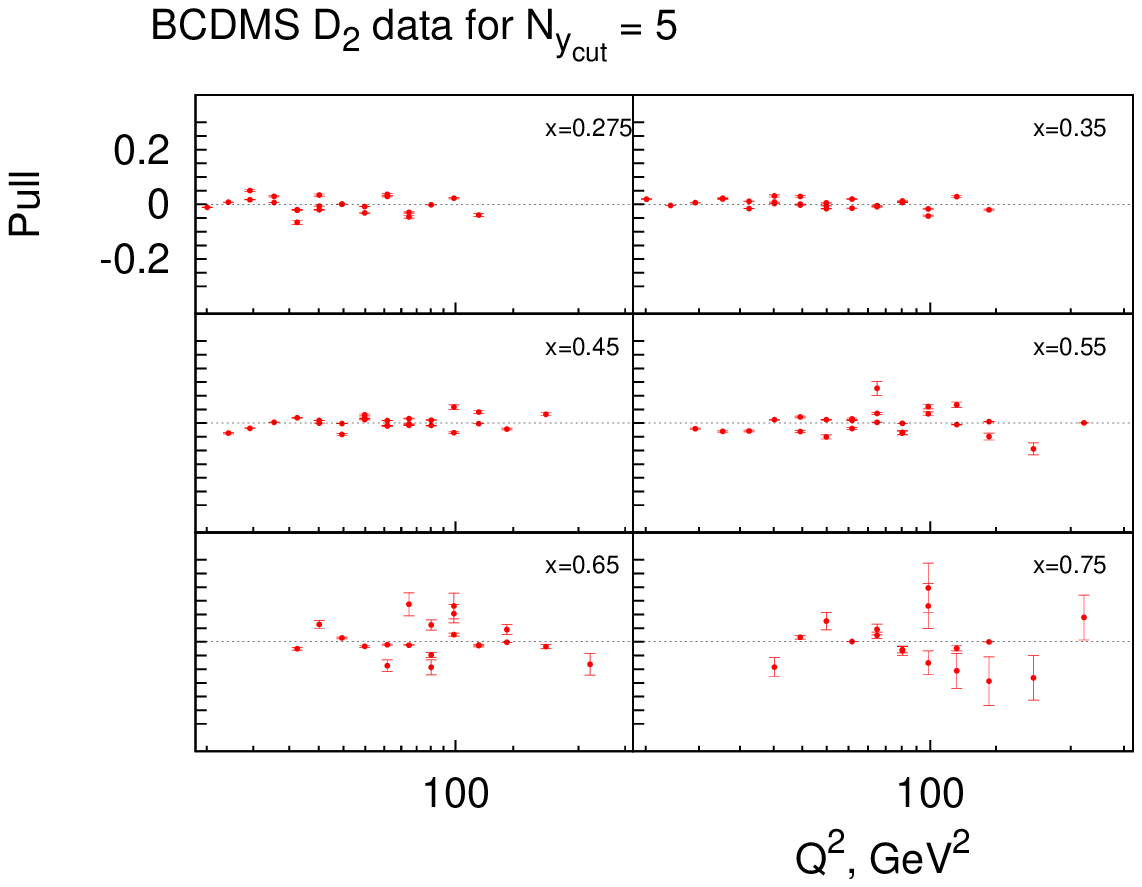,width=150mm,height=95mm}
}
\end{picture}
\vskip 0.2cm
\caption{ The same as in Fig.~1 for BCDMS deuterium data.}
\end{figure}

\begin{figure}[!htb]
\unitlength=1mm
\begin{picture}(0,100)
  \put(0,-5){%
   \psfig{file=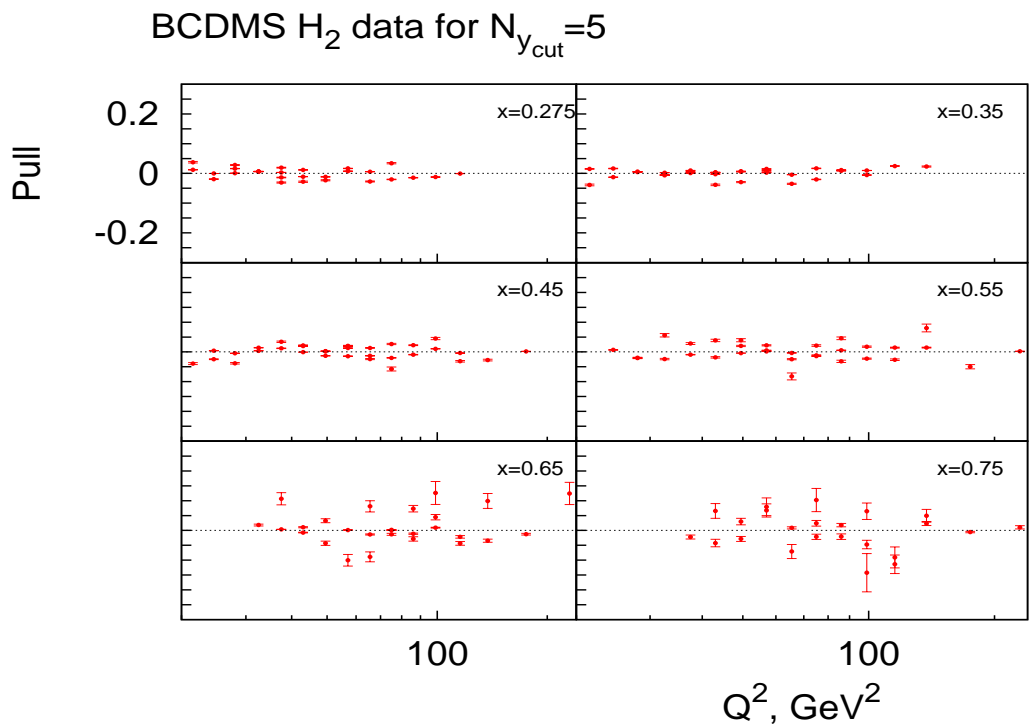,width=150mm,height=95mm}
}
\end{picture}
\vskip 0.2cm
\caption{ The same as in Fig.~1 for BCDMS hydrogen data.}
\end{figure}

\begin{figure}[!htb]
\unitlength=1mm
\begin{picture}(0,100)
  \put(0,-5){%
   \psfig{file=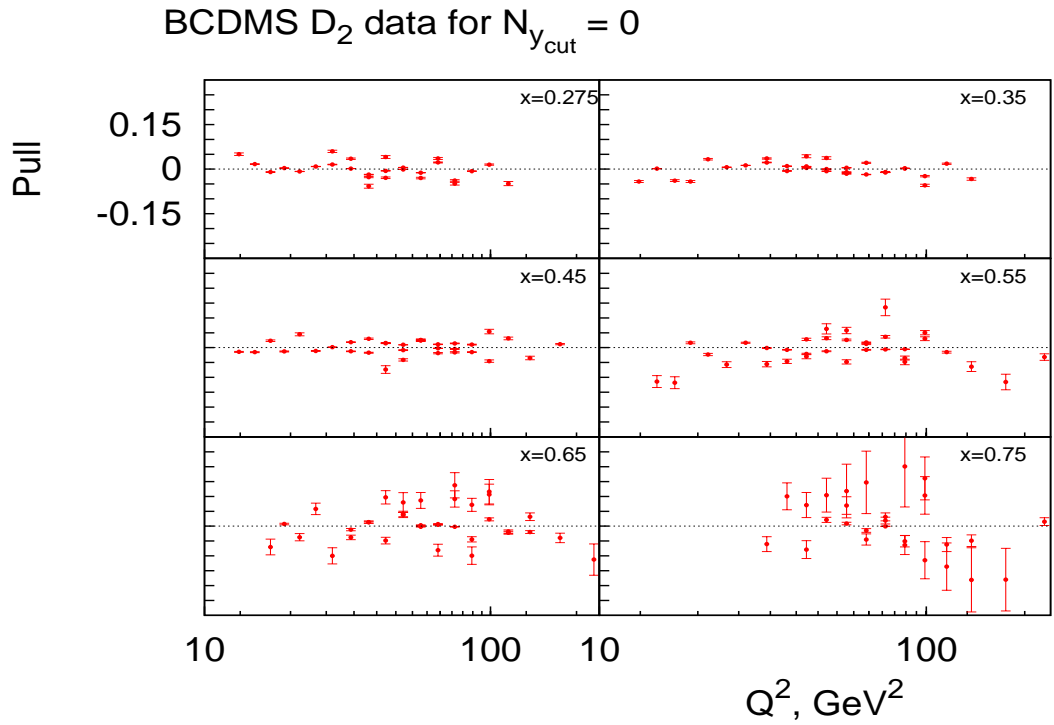,width=150mm,height=95mm}
}
\end{picture}
\vskip 0.2cm
\caption{ The same as in Fig.~1 for BCDMS deuterium data before $y$ cuts are imposed.}
\end{figure}

\begin{figure}[!htb]
\unitlength=1mm
\begin{picture}(0,100)
  \put(0,-5){%
   \psfig{file=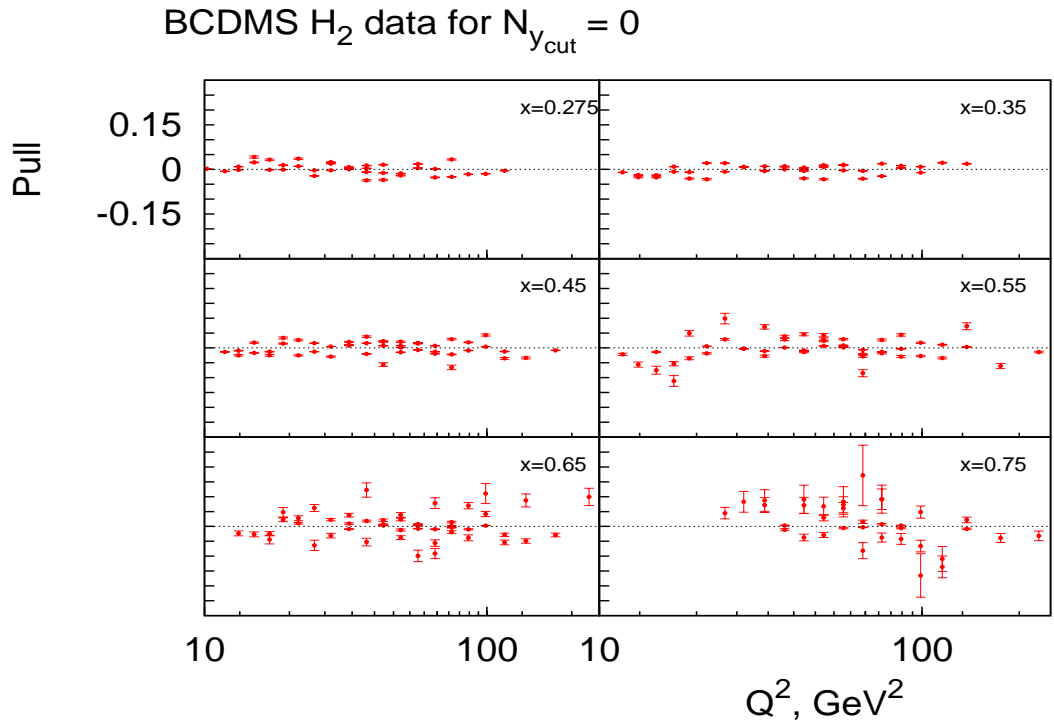,width=150mm,height=95mm}
}
\end{picture}
\vskip 0.2cm
\caption{ The same as in Fig.~1 for BCDMS hydrogen data before $y$ cuts are imposed.}
\end{figure}


\end{document}